\documentclass[sigconf,screen,9pt]{acmart}

\renewcommand\footnotetextcopyrightpermission[1]{} % removes the permission block
\usepackage{tikz}
\newcommand*\circled[1]{\tikz[baseline=(char.base)]{
            \node[shape=circle,draw,inner sep=2pt] (char) {#1};}}

\usepackage{amsmath,amssymb,amsfonts}
\usepackage{braket}
\usepackage{booktabs}
\usepackage{algorithmic}
\usepackage{graphicx}
\usepackage[most]{tcolorbox}
\usepackage{textcomp}
\usepackage{bmpsize}
\usepackage{xcolor}
\usepackage[linesnumbered,ruled]{algorithm2e} 
\usepackage{algorithmic}
\usepackage{lipsum}
\usepackage{filecontents}
\usepackage{caption}
\usepackage{subcaption}
\usepackage{adjustbox}
\usepackage{setspace}
\usepackage{tabularx}
\usepackage{balance}
\usepackage{pifont}
\usepackage{bbding}
\usepackage{multicol,multirow}
\usepackage{color}
\usepackage{soul}
\tcbset{
  mytextbox/.style={
    colback=blue!10!white,      % background color
    colframe=blue!30!white, % frame color
    fonttitle=\bfseries,
    coltitle=black,
    title={Key Takeaways:},
    boxrule=0.8pt,
    arc=4pt,
    left=6pt, right=6pt, top=6pt, bottom=6pt
  }
}


\author{Navnil Choudhury}
\affiliation{%
  \institution{Rensselaer Polytechnic Institute}
  \city{Troy}
  \country{USA}
}
\email{choudn3@rpi.edu}

\author{Yizhuo Tan}
\affiliation{%
  \institution{Yale University}
  \city{New Haven}
  \country{USA}
}
\email{yizhuo.tan@yale.edu}

\author{Jiaqi Yu}
\affiliation{%
  \institution{Northwestern University}
  \city{Illinois}
  \country{USA}
}
\email{jiaqi.yu1@northwestern.edu}

\author{Jakub Szefer}
\affiliation{%
  \institution{Northwestern University}
  \city{Illinois}
  \country{USA}
}
\email{jakub.szefer@northwestern.edu}

\author{Kanad Basu}
\affiliation{%
  \institution{Rensselaer Polytechnic Institute}
  \city{Troy}
  \country{USA}
}
\email{basuk@rpi.edu}

\begin{document}

\title{EPAR: Electromagnetic Pathways to Architectural Reliability in Quantum Processors}

\begin{abstract}
% As superconducting quantum processors scale, understanding how physical layout shapes qubit interactions has become essential for architectural reliability. However, existing approaches offer limited visibility into how layout-dependent electromagnetic behavior translates into quantum execution-level effects. In this paper, we present EPAR, a unified electromagnetic-to-architecture framework revealing how field-driven disturbances manifest as connectivity distortions within realistic superconducting layouts. 
% % By coupling electromagnetic modeling with architectural interpretation, EPAR provides a principled basis for improving layout robustness, demonstrating up to 15\% reductions in connectivity distortion and up to 4$\times$ reductions in directional vulnerability. EPAR’s predictions are further validated on observed sensitivity patterns in Rigetti Ankaa-3, underscoring its architectural relevance.
% By unifying electromagnetic modeling with architectural analysis, EPAR uncovers quantitative robustness regimes, showing stable behavior for \textbf{LTD $< 0.15$} and \textbf{SI $< 0.10$}, pulse-dependent fragility for intermediate distortion, and \textbf{sharp fidelity collapse} beyond \textbf{LTD $> 0.80$} or \textbf{SI $> 0.25$}. EPAR also demonstrates that edges with identical two-qubit error rates can differ by \textbf{more than $10\times$} in dynamic robustness.

As superconducting processors scale, understanding how physical layout shapes qubit interactions is essential for architectural reliability. Existing methods offer limited insight into how electromagnetic design choices translate into execution-level behavior. We present EPAR, an electromagnetic-to-architecture framework that predicts robustness early directly from physical design by reconstructing how design distortion modifies the effective Hamiltonian, reroutes mediated connectivity, and influences control-pulse response. Across all tested layouts, EPAR’s structural scores show 100\% agreement with two-qubit error trends yet reveal over 10$\times$ robustness differences among edges with identical calibrated error rates, going beyond conventional metrics to provide improved and actionable compiler guidance.

\end{abstract}
\maketitle

\pagestyle{plain}

\section{Introduction} \label{sec:intro}

Quantum computing promises computational capabilities that surpass classical systems by leveraging quantum superposition, entanglement, and interference~\cite{nielsen_and_chuang, grover1996fast}. Among existing technologies, superconducting qubits have emerged as one of the leading platforms for building near-term and intermediate-scale processors, driven by rapid advances in fabrication, device control, and integrated software stacks~\cite{photonic, trappedion, superconducting, superconducting2}. As superconducting processors grow in size and complexity, understanding how the physical layout of qubits, couplers, and interconnect structures shape computational behavior has become increasingly important~\cite{study1, study2}. However, most architecture-level and compiler toolchains still rely on an idealized, vendor-specified coupling graph, modeling variability only through gate error rates and coherence metrics~\cite{qiskit1, murali2020software}. This abstraction hides layout-dependent electromagnetic interactions that extend beyond the nominal topology, limiting visibility into the physical processes that govern multi-qubit coupling~\cite{crosstalk}.

This disconnect between nominal and actual topology arises because real superconducting layouts introduce parasitic couplings and subtle geometric asymmetries that reshape the underlying interaction landscape~\cite{geller2015tunable}. Small variations in conductor geometry or island placement can alter coupling strengths, generate unintended mediated pathways, and shift qubit and coupler frequencies 
% and produce direction-dependent crosstalk 
even in ostensibly symmetric layouts~\cite{geometry}.
Together, these mechanisms cause the hardware’s actual interaction topology to deviate from the logical topology assumed during design. This mismatch can heavily affect gate behavior and multi-qubit interactions upon starting system operation~\cite{kandala2017hardware}.
% Collectively, these mechanisms cause the effective interaction topology realized by the hardware to differ meaningfully from the logical topology assumed during design, influencing gate behavior and multi-qubit interactions at the earliest stages of system operation.

\begin{figure}[t]
% \vspace{-4mm}
\centering
  \includegraphics[width=0.8\linewidth]{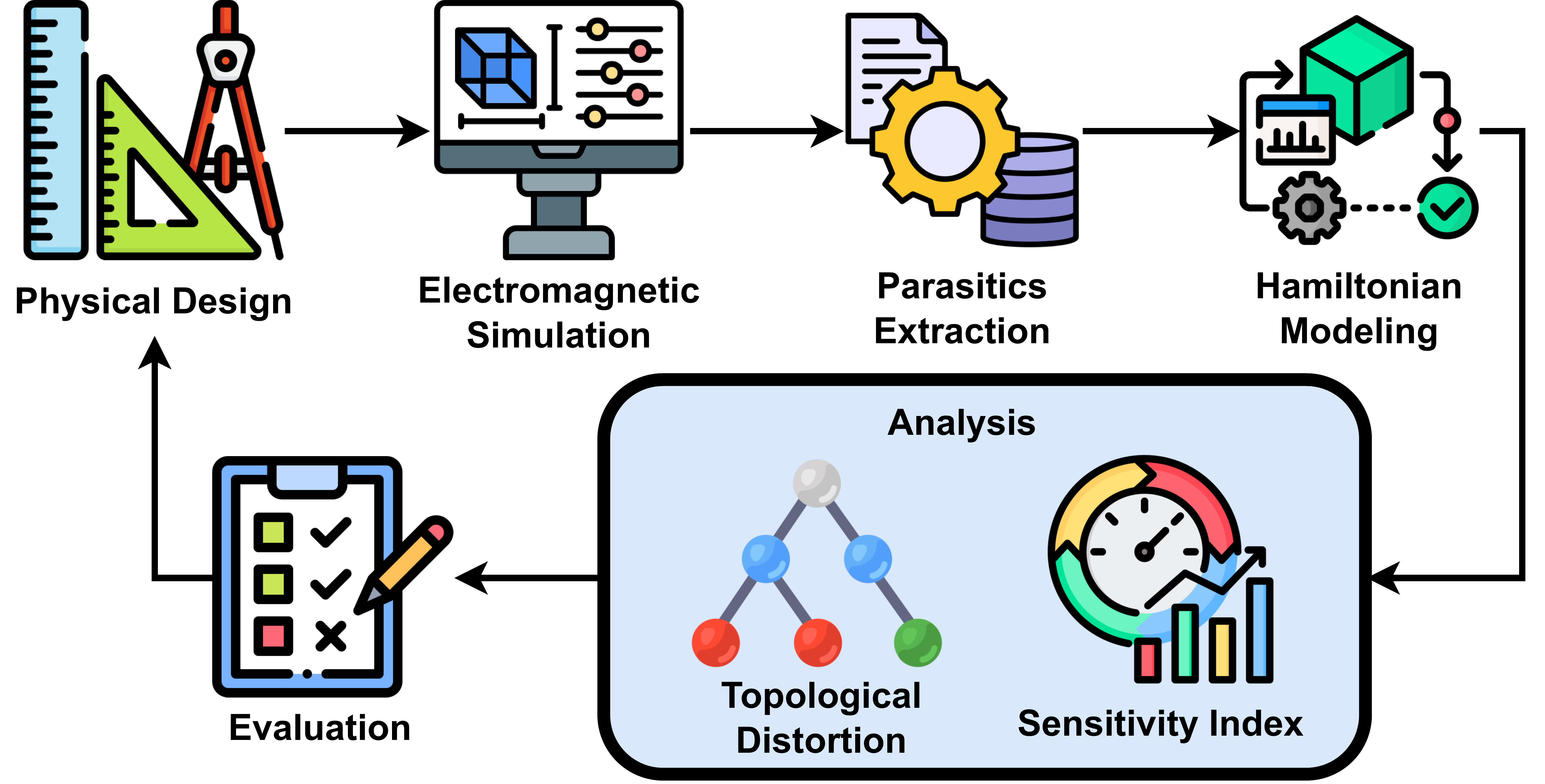}
  % \vspace{-5mm}
  \caption{Overview of our proposed EPAR workflow.}
  \label{fig:epar_intro}
  \vspace{-5mm}
\end{figure}

Despite the architectural significance of these layout-driven effects, existing software tools offer little insight into how geometry-dependent interactions arise or propagate through the system~\cite{geller2015tunable}. Although existing calibration pipelines report essential gate and coherence properties, they do not indicate how the physical layout shaped the structure of the underlying interactions. Architecture-level frameworks and circuit-level simulators inherit the same limitation: they operate on fixed nominal coupling maps, modeling variability through noise parameters rather than through the electromagnetic processes that determine actual coupling strengths~\cite{tket, openql, qutip, qulacs}. As a result, the interaction distortions introduced by layout, whether parasitic couplings, geometric asymmetries, or mediated pathways, are not considered in conventional workflows, despite their influence on mapping, scheduling, and routing decisions.

In this paper, we introduce EPAR, a novel and transformative layout-aware analysis framework that derives the effective interaction structure of a superconducting processor directly from its physical geometry. Unlike existing toolflows that treat electromagnetic modeling and architectural
analysis as disjoint stages, EPAR unifies electromagnetic simulation,
capacitance extraction, Hamiltonian reconstruction, and time-domain evaluation
into a single physics-grounded pipeline. This integration provides unprecedented
visibility into how parasitic couplings, mediated pathways, and geometric asymmetries originate from the layout and reshape the effective interaction
topology. The overall workflow is summarized in Figure~\ref{fig:epar_intro}. As shown in the figure, EPAR begins with the physical layout of a superconducting processor and
applies a full electromagnetic simulation to extract geometry-dependent
parasitics. These parasitics are used to reconstruct the multi-mode Hamiltonian. From this, EPAR derives two system-level metrics—Logical Topology Distortion (LTD) and Sensitivity Index (SI)—which quantify how the effective connectivity and pulse response deviate from the intended design. This geometry-derived interaction landscape enables principled refinement of the physical layout and informed compiler decisions for mapping, routing, and pulse-schedule selection.

\textit{By elevating layout-driven physics to the architectural and compiler levels, EPAR provides the first systematic method to evaluate how geometry influences qubit-neighbor relationships, coupler robustness, parallelism limits, and routing strategies. This enables a new class of design-time analyses that guide placement decisions, assess topology reliability, and support the development of more physically faithful compilation heuristics.
While full-chip EM simulation is infeasible at scale, EPAR remains tractable by modeling only the layout-defined capacitance graph and analytically reconstructing the effective interaction topology.}
Our main contributions are as follows:

\begin{itemize}

\item This paper presents EPAR, a unified pipeline that combines electromagnetic simulation, capacitance extraction, Hamiltonian reconstruction, and interaction-graph inference. Together, these stages enable architectural analysis grounded in the device’s real physics rather than an idealized vendor coupling map.

% \item We identify how layout geometry disrupts intended logical topology by uncovering hidden couplings, suppressed connections, and asymmetric interaction strengths that reshape qubit–neighbor relationships and compromise routing and placement assumptions used in existing toolchains.

\item We identify how layout geometry deforms intended logical connectivity by revealing hidden couplings and asymmetric interaction strengths that undermine routing and placement assumptions in existing toolchains. To this end, we introduce Logical Topology Distortion (LTD) to capture these layout-induced effects quantitatively.

\item We capture how local parameter shifts propagate across the processor, exposing leakage paths and direction-dependent vulnerabilities shaped by capacitive-island layout and geometric asymmetries. To this end, we introduce the Sensitivity Index (SI) to capture these behaviors.

% \item We determine how localized parameter variations propagate across the processor, exposing orientation-dependent crosstalk channels and directional vulnerabilities caused by capacitive island placement and geometric asymmetries that influence qubit placement and parallel-gate execution.

\item Our evaluations show that LTD and SI uncover layout-induced instabilities that are invisible to calibrated 2Q error rates. When LTD $<0.15$, all pulse families maintain high fidelity. Moderate distortion ($0.3<$LTD$<0.6$) introduces geometry- and pulse-dependent errors, while severe distortion (LTD $>0.8$) drives infidelity toward $0.8$–$1.0$. SI reveals regime-dependent failure patterns: high-SI layouts remain below $10\%$ error in short/medium regimes, whereas mid-SI and low-SI layouts collapse ($\sim 60$–$96\%$) under longer or overdrive pulses, providing actionable compiler guidance.

\end{itemize}

\section{Background}

\subsection{Transmon Qubits}
Superconducting quantum processors are among the leading platforms for building scalable quantum computers~\cite{superconducting}. The discovery of superconducting qubits like the transmon, fluxonium, bosonic, and more complex circuits, facilitated the advancements of coupling between qubits has paved the way for rapid evolution of superconductor-based quantum computers~\cite{transmon, fluxonium}. 
As the most widely used superconducting qubit architecture, the transmon qubit is built from a Josephson Junction (JJ), which contains two superconducting electrodes that are separated by a thin insulating barrier, allowing for the coherent tunneling of Cooper pairs, and a shunting capacitor that suppresses sensitivity to charge noise~\cite{josephson}. Together, these components form a nonlinear LC oscillator whos quantum energy encode the qubit, and its effective Hamiltonian is given by 
% as shown in Fig~\ref{}, whose effective Hamiltonian is given by
\begin{align}
    H = 4 E_C n^2 - E_J cos(\varphi)
\end{align}
where $E_C$ denotes the capacitive charging energy, $n$ is the number of excess Cooper pairs, $E_J$ is the Josephson energy, and $\varphi$ is the phase difference across the Josephson junction, whose operator satisfies $[\hat{\varphi}, \hat{n}]=i$. In the regime $E_C \ll E_J$, the qubit becomes less sensitive to charge noise, making it possible towards high coherence.
Single-qubit gates are implemented using shaped microwave pulses that drive transitions between $\ket{0}$ and $\ket{1}$, while two-qubit gates exploit controlled interactions between neighboring qubits, such as cross-resonance or tunable-frequency exchange.

\begin{figure}[t]
% \vspace{-4mm}
\centering
  \includegraphics[width=0.7\linewidth]{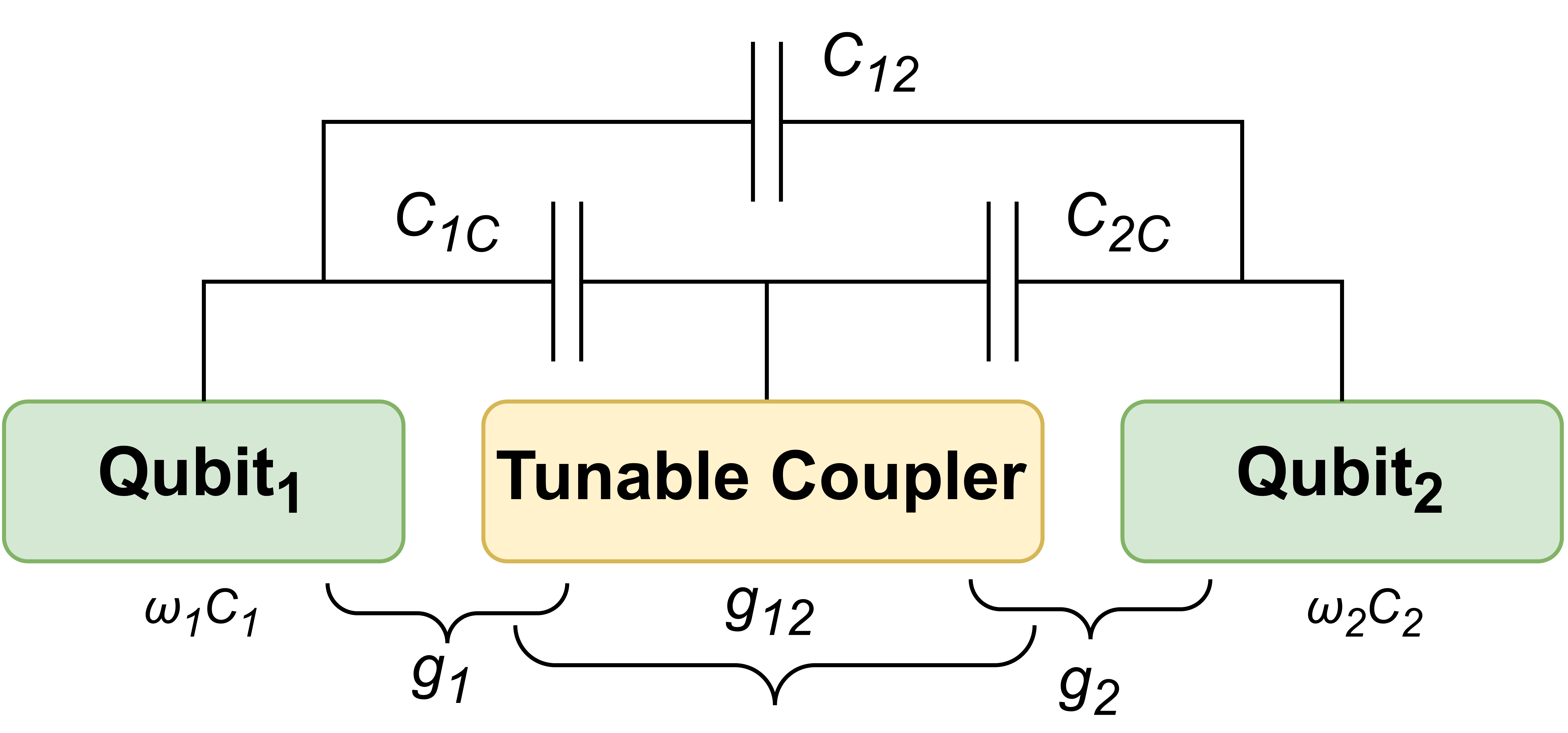}
  % \vspace{-5mm}
  \caption{Circuit-level view of a two-qubit one coupler subsystem.}
  \label{fig:coupler}
  \vspace{-5mm}
\end{figure}

\subsection{Tunable Couplers}
Multi-qubit gates rely on precisely controlled interactions between transmons, and tunable couplers enable this by dynamically adjusting the effective qubit–qubit coupling, avoiding the residual interactions inherent to fixed layouts~\cite{geller2015tunable, blais2003tunable, yan2018tunable}. In the standard three-mode setup (Fig.~\ref{fig:coupler}), each qubit couples to a central tunable element; applying a Schrieffer–Wolff reduction yields the effective two-qubit Hamiltonian
\[
\tilde{H}\;=\;\tfrac{1}{2}\tilde{\omega}_1\sigma_1^z+\tfrac{1}{2}\tilde{\omega}_2\sigma_2^z+\tilde{g}(\sigma_1^+\sigma_2^-+\sigma_2^+\sigma_1^-),
\]
where the interaction rate $\tilde{g}\!\propto\!g_1 g_2/\Delta + g_{12}$ can be tuned on for entangling operations and near-off during idle periods via precisely shaped microwave pulses and the coupler frequency.

\section{Proposed EPAR Workflow} \label{sec:method}

\noindent We develop EPAR to quantify how layout-induced electromagnetic effects reshape both the effective logical topology and the operational behavior of a superconducting processor, enabling accurate prediction and mitigation of architectural impacts. EPAR follows a three-stage workflow linking physical layout, electromagnetic interactions, and system-level behavior.
\circled{1} First, EPAR derives a parasitic-aware capacitance matrix and the corresponding effective Hamiltonian by modeling the electromagnetic interactions from the physical layout (Section~\ref{subsec:geom_to_hamiltonian}).
Building on this Hamiltonian, \circled{2} \textbf{Logical Topology Distortion (LTD)} is introduced to capture how the realized interaction graph deviates from the nominal design topology, resulting in compromised routing paths, and reduced fidelity of connectivity-driven optimizations (Section~\ref{subsec:ltd}).
\circled{3} We further define the \textbf{Sensitivity Index (SI)} to capture how local parameter variations propagate to non-target qubits and resonators, guiding mapping and scheduling decisions(Section~\ref{subsec:csi}).

\subsection{Graph-Based Hamiltonian Reconstruction from Layout}
\label{subsec:geom_to_hamiltonian}

EPAR introduces a layout-driven reconstruction pipeline in which the physical geometry is treated as an interaction graph rather than an idealized circuit. Each conducting body (qubit pads, coupler islands, ground structures) becomes a node, and the full geometry-derived capacitance matrix $\mathbf{C}$ defines weighted edges that include not only the intended couplings but also the weak parasitic pathways introduced by spatial proximity and asymmetries. This yields a geometry-specific interaction graph $G_{\mathrm{geom}}=(V,E,\mathbf{C})$ that explicitly encodes all electrostatic channels implied by the layout—structure that is discarded in nominal circuit models.

From this graph, we construct the charging-energy matrix
\begin{equation}
    \mathbf{E}_C = \frac{(2e)^2}{2}\,\mathbf{C}^{-1},
\end{equation}
and incorporate the Josephson potentials $U_{J,i}(\phi_i)$ for nonlinear nodes to obtain the effective Hamiltonian
\begin{equation} \label{eq:Heff}
H_{\mathrm{eff}} = 4\,\mathbf{n}^{\mathsf{T}}\mathbf{E}_C\,\mathbf{n} + \sum_{i\in\mathcal{J}} U_{J,i}(\phi_i).
\end{equation}
This graph-based reconstruction is inherently layout-specific: geometric distortions, mediated interactions, and unintended coupling channels emerge directly from $\mathbf{E}_C$, enabling downstream LTD and SI metrics to analyze how physical layout reshapes qubit interactions in a way that traditional schematic-based models cannot. 
% Our approach follows standard practice for tunable-coupler transmons, where capacitive pathways dominate the static interaction structure~\cite{geller2015tunable}.

% We then use this Hamiltonian to reconstruct the effective physical connectivity graph implied by the device’s electromagnetic behavior and compare it with the nominal design topology.

% \subsection{Logical Topology Distortion (LTD)}
% \label{sec:ltd}
% LTD measures the extent to which layout-dependent electromagnetic structure alters the 
% intended connectivity.  
% It captures both missing intended edges (suppressed by geometry) and parasitic edges 
% (introduced purely by layout), as well as asymmetric distortions of coupling strengths 
% that may influence qubit-neighbor suitability and routing reliability.  
% The procedure is formalized in Algorithm~\ref{alg:ltd}.

\subsection{Logical Topology Distortion (LTD) Reconstruction}
\label{subsec:ltd}

Following the generation of the effective Hamiltonian in the previous subsection, we now use it to reconstruct the interaction topology implied by the device geometry and assess how it differs from the nominal architectural specification. Our objective in this stage is to determine how capacitive layout features modify the effective connectivity graph that governs qubit behavior. To this end, we propose \emph{Algorithm~\ref{alg:ltd_reconstruction}} to recover
the physical interaction graph directly from the Hamiltonian and compute the
Logical Topology Distortion (LTD), which captures how the geometry-induced
connectivity deviates from the nominal architectural topology. Formally, we
define
\begin{equation}
    \mathrm{LTD}
    = \frac{\,|E^{+}| + |E^{-}|\,}{|E_{\mathrm{nom}}|}
\end{equation}
where $E^{+} = E_{\mathrm{phys}} \setminus E_{\mathrm{nom}}$ denotes parasitic
edges introduced by the layout, and $E^{-} = E_{\mathrm{nom}} \setminus
E_{\mathrm{phys}}$ denotes intended edges weakened or removed by geometric
asymmetries. The algorithm operates on three inputs: the nominal design graph $G_{\mathrm{nom}} = (V, E_{\mathrm{nom}})$ which specifies the intended qubit connectivity, the geometry-derived Hamiltonian $H_{\mathrm{eff}}$, and a significance threshold $\tau_{\min}$.
% Using these quantities, \emph{Algorithm~\ref{alg:ltd_reconstruction}} reconstructs the physical interaction graph directly from the Hamiltonian and computes a quantitative measure of logical topology distortion.

\emph{Algorithm~\ref{alg:ltd_reconstruction}} begins by extracting the pairwise interaction terms $g_{ij}$ embedded within $H_{\mathrm{eff}}$, from Equation~\ref{eq:Heff} (\textit{line~1}). These coefficients encode both the intended qubit--coupler interactions and the parasitic cross-capacitive effects introduced by the geometry. To ensure that subsequent decisions depend only on the relative strength of these interactions, we normalize the couplings (\textit{line~2}) according to:
\begin{equation}
    \tilde{g}_{ij} = \frac{g_{ij}}{\max |g_{ij}|}
\end{equation} 
All pairs $(i,j)$ with nonzero normalized interaction strength $\tilde{g}_{ij}$ extracted from the effective Hamiltonian are collected into the candidate edge set $E_{\mathrm{cand}}$, representing all layout-induced couplings prior to filtering. Applying the threshold $\tau_{\min}$ to these interactions yields the effective physical edge set, which captures the subset of couplings that are physically realized in the hardware, as follows:
\begin{equation}
    E_{\mathrm{phys}} = \{(i,j) \in E_{\mathrm{cand}} \;|\; \tilde{g}_{ij} \ge \tau_{\min}\},
\end{equation}
The resulting $E_{\mathrm{cand}}$ is used to define the reconstructed physical graph $G_{\mathrm{phys}}$, with edge weights given by the normalized coupling strengths (\textit{line~5}).
We then compare $G_{\mathrm{phys}}$ against the nominal design graph (defined in Section 3.1) to identify discrepancies arising from electromagnetic behavior. Edges that appear in the nominal architecture $E_{\mathrm{nom}}$ but not in the reconstructed topology $E_{\mathrm{phys}}$ are recorded as missing interactions given by:
\begin{equation}
    E^{-} = E_{\mathrm{nom}} \setminus E_{\mathrm{phys}},
\end{equation}
On the other hand, edges that emerge solely due to parasitic coupling pathways are captured as shown below (\textit{lines~6-7}):
\begin{equation}
    E^{+} = E_{\mathrm{phys}} \setminus E_{\mathrm{nom}},
\end{equation}
For every interacting pair, we additionally compute a directional asymmetry value given by:
\begin{equation}
    A_{ij} = |\tilde{g}_{ij} - \tilde{g}_{ji}|,
\end{equation}
This allows us to evaluate whether geometric imbalance or ground-return effects produce coupling that is stronger in one direction than the other (\textit{line~8}). These values are then aggregated into a global asymmetry profile (\textit{line~9}).

Following this, to quantify the overall degree of distortion in the physical layout, we compute the Logical Topology Distortion (LTD) metric (\textit{line~10}), defined as:
\begin{equation}
    \mathrm{LTD} = \frac{|E^{+}| + |E^{-}|}{|E_{\mathrm{nom}}|},
\end{equation}
 This measure reflects the proportion of edges that either fail to manifest physically or arise solely due to parasitic interactions. We also compute a severity-weighted variant as follows:
\begin{equation}
    \mathrm{LTD}_{\mathrm{w}} =
    \frac{\sum_{(i,j)\in E^{+}} \tilde{g}_{ij} \;+\; \sum_{(i,j)\in E^{-}} 1}
    {\sum_{(i,j)\in E_{\mathrm{nom}}} 1},
\end{equation}
This incorporates the magnitude of parasitic couplings and distinguishes mild structural deviation from severe topology reshaping (\textit{line~11}).

\begin{algorithm}[t]
\raggedright
\caption{Logical Topology Distortion (LTD) Reconstruction}
\label{alg:ltd_reconstruction}
{\footnotesize
\KwIn{$G_{\mathrm{nom}} = (V, E_{\mathrm{nom}})$; $H_{\mathrm{eff}}$; threshold $\tau_{\min}$}
\KwOut{$G_{\mathrm{phys}}$; LTD; $E^{+}$; $E^{-}$; $A_{ij}$}
\setcounter{AlgoLine}{0}
\SetAlgoLined

Extract $g_{ij}$ from $H_{\mathrm{eff}}$\;

Normalize $\tilde{g}_{ij} \leftarrow g_{ij} / \max |g_{ij}|$\;

$E_{\mathrm{cand}} \leftarrow \{(i,j) \,|\, \tilde{g}_{ij} > 0\}$\;

$E_{\mathrm{phys}} \leftarrow \{(i,j) \in E_{\mathrm{cand}} : \tilde{g}_{ij} \ge \tau_{\min}\}$\;

$G_{\mathrm{phys}} \leftarrow (V, E_{\mathrm{phys}}, \tilde{g}_{ij})$\;

$E^{-} \leftarrow E_{\mathrm{nom}} \setminus E_{\mathrm{phys}}$\;

$E^{+} \leftarrow E_{\mathrm{phys}} \setminus E_{\mathrm{nom}}$\;

$A_{ij} \leftarrow |\tilde{g}_{ij} - \tilde{g}_{ji}|$\;

Compute global asymmetry profile\;

$\mathrm{LTD} \leftarrow \dfrac{|E^{+}| + |E^{-}|}{|E_{\mathrm{nom}}|}$\;

$\mathrm{LTD}_{\mathrm{w}} \leftarrow 
\dfrac{\sum_{(i,j)\in E^{+}} \tilde{g}_{ij} + \sum_{(i,j)\in E^{-}} 1}
      {\sum_{(i,j)\in E_{\mathrm{nom}}} 1}$\;

Summarize deviations\;

\Return{$G_{\mathrm{phys}}, \mathrm{LTD}, E^{+}, E^{-}, A_{ij}$}\;
}
\end{algorithm}

Finally, Algorithm~\ref{alg:ltd_reconstruction} returns the reconstructed physical graph along with the LTD metrics, the parasitic and missing edge sets, and the directional asymmetry profile (\textit{lines~12–13}). Each of these outputs serves a distinct role: \textit{First}, the physical graph captures the geometry-implied interaction structure. \textit{Next} The parasitic and missing edges identify where layout effects introduce or suppress couplings relative to the design intent. \textit{Finally}, the asymmetry profile reveals non-reciprocal interaction strengths that can affect routing and control.
Together, these parameters provide an actionable characterization of how geometric features reshape the effective interaction topology, enabling downstream architectural tools and compilers to reason about, anticipate, and correct for layout-induced distortions rather than operating under incorrect connectivity assumptions.

\subsection{Sensitivity Index (SI) Reconstruction}
\label{subsec:csi}

Following the reconstruction of the physical interaction topology using the LTD metric, we now evaluate its susceptibility to localized perturbations in the system’s Hamiltonian parameters. Our objective in this stage is to determine how small fluctuations at a single node influence parasitic interaction channels throughout the layout-induced coupling network. To this end, we introduce the \emph{Sensitivity Index} (SI), which quantifies how
strongly a perturbation applied to one node propagates across the system and
modifies unintended couplings. Formally, for a perturbation $\delta$ applied at
node $k$, SI is defined as
\begin{equation}
    \mathrm{SI}(k)
    = \max_{i,j} \left| 
        \frac{ g_{ij}^{(k,\delta)} - g_{ij} }{\delta}
    \right|
\end{equation}
where $g_{ij}$ denotes the unperturbed interaction strength between nodes $i$ and
$j$, and $g_{ij}^{(k,\delta)}$ denotes the interaction strength after
perturbing node $k$ by $\delta$. 
% This analysis identifies geometrically vulnerable nodes—those whose fluctuations disproportionately reshape parasitic pathways—and provides a physics-grounded indicator of dynamical crosstalk risk.

Given the effective Hamiltonian $H_{\mathrm{eff}}$ and the reconstructed physical graph $G_{\mathrm{phys}}$, we proceed by using \emph{Algorithm~\ref{alg:csi_reconstruction}} to compute the Sensitivity Index (SI). We begin by introducing a small perturbation $\delta$ at each node $k$ and evaluating how this perturbation modifies the system’s interaction structure. We first extract the baseline pairwise interaction strengths $g_{ij}$ from $H_{\mathrm{eff}}$, in Equation~\ref{eq:Heff}, and normalize them using $\tilde{g}{ij} = g{ij} / \max |g_{ij}|$, establishing a consistent reference for comparing perturbed values (\textit{lines~1-2}).
Subsequently, we identify the parasitic edge set given by:
\begin{equation}
    E^{+} = E_{\mathrm{phys}} \setminus E_{\mathrm{nom}}
\end{equation}
which isolates all couplings that appear in the geometry-derived graph but are absent from the nominal design, thereby marking the unintended interaction channels on which SI is evaluated (\textit{line~3}).

\begin{algorithm}[t]
\raggedright
\caption{Sensitivity Index (SI) Reconstruction}
\label{alg:csi_reconstruction}
{\footnotesize
\KwIn{$H_{\mathrm{eff}}$; $G_{\mathrm{phys}} = (V, E_{\mathrm{phys}})$; $G_{\mathrm{nom}} = (V, E_{\mathrm{nom}})$; perturbation magnitude $\delta$}
\KwOut{$\mathrm{SI}(k)$; $\Delta g_{ij}^{(k)}$; $S_{k \rightarrow (i,j)}$}
\setcounter{AlgoLine}{0}
\SetAlgoLined

Extract baseline pairwise interactions $g_{ij}$ from $H_{\mathrm{eff}}$\;

Normalize baseline interactions:
$\tilde{g}_{ij} = g_{ij} / \max |g_{ij}|$\;

Identify parasitic edges:
$E^{p} = E_{\mathrm{phys}} \setminus E_{\mathrm{nom}}$\;

\For{each node $k \in V$}{
    Compute local sensitivity operator $\Delta_k$\;

    Form perturbed Hamiltonian:
    $H_{\mathrm{eff}}^{(k,\delta)} = H_{\mathrm{eff}} + \delta \cdot \Delta_k$\;

    Extract perturbed interactions $g_{ij}^{(k,\delta)}$ from $H_{\mathrm{eff}}^{(k,\delta)}$\;

    Normalize perturbed interactions using the same scale:
    $\tilde{g}_{ij}^{(k)} = g_{ij}^{(k,\delta)} / \max |g_{ij}|$\;

    Compute deviations:
    $\Delta g_{ij}^{(k)} = \tilde{g}_{ij}^{(k)} - \tilde{g}_{ij}$\;

    \For{each $(i,j) \in E^{+}$}{
        $S_{k \rightarrow (i,j)} = |\Delta g_{ij}^{(k)}|$\;
    }

    Compute node-level Sensitivity Index:
    $\mathrm{SI}(k) = \sum_{(i,j)\in E^{p}} S_{k \rightarrow (i,j)}$\;
}

Assemble vector $(\mathrm{SI}(0), \mathrm{SI}(1), \ldots)$\;

\Return{$\mathrm{SI}(k)$, $\Delta g_{ij}^{(k)}$, $S_{k \rightarrow (i,j)}$}\;
}
\end{algorithm}

The core of the analysis occurs inside the per-node loop (\textit{lines~4-14}), where we explicitly evaluate how a perturbation at each individual node reshapes the interaction landscape and propagates through the parasitic coupling network.
For each node $k$, we construct the local sensitivity operator $\Delta_k$, which specifies the direction in Hamiltonian parameter space corresponding to a localized perturbation at that node (\textit{lines~4-5}). 
% This operator ensures that the applied displacement $\delta$ modifies only the terms associated with node $k$ and cleanly isolates its contribution to the resulting change in the interaction structure.
Following this, for eah node $k$, we construct the perturbed Hamiltonian as:
\begin{equation}
H_{\mathrm{eff}}^{(k,\delta)} = H_{\mathrm{eff}} + \delta \cdot \Delta_k
\end{equation}
Our approach ensures that the perturbation affects only the degrees of freedom tied to that specific node (\textit{lines~6}). From this perturbed model, we extract the modified interaction strengths $g_{ij}^{(k,\delta)}$, which capture how the perturbation at $k$ alters each pairwise coupling (\textit{line~7}). We then normalize these perturbed values using the same scale factor as the baseline interactions (\textit{line~8}), yielding $\tilde{g}_{ij}^{(k)}$. This shared normalization guarantees that all deviations are measured on a consistent scale and reflect relative changes in the coupling strengths.
Following this, we compute the deviation in Equation~\ref{eq:gij}, (\textit{line~9}), which quantifies how strongly perturbing node $k$ alters each interaction channel $(i,j)$ and thus reveals that node’s influence on system-wide parasitic couplings.
\begin{equation}
\Delta g{ij}^{(k)} = \tilde{g}{ij}^{(k)} - \tilde{g}{ij}
\label{eq:gij}
\end{equation}
To focus specifically on geometry-induced vulnerabilities, we further isolate the parasitic edge set in Equation~\ref{eq:ep} (\textit{line~10}), since these couplings do not exist in the nominal design and therefore represent unintended interaction pathways created by the layout. 
\begin{equation}
E^{p} = E_{\mathrm{phys}} \setminus E_{\mathrm{nom}},
\label{eq:ep}
\end{equation}
 For every parasitic edge $(i,j)\in E^{p}$, we measure the influence of node $k$ using the sensitivity term, as defined in Equation~\ref{eq:sk} (\textit{lines~11-12}), which isolates how strongly node $k$ modulates that specific unintended coupling.
\begin{equation}
S_{k \rightarrow (i,j)} = \left| \Delta g_{ij}^{(k)} \right|
\label{eq:sk}
\end{equation}
 Aggregating these contributions yields the node-level \emph{Sensitivity Index}, in Equation~\ref{eq:sik} (\textit{lines~13-14}), providing a scalar measure of how perturbations originating at node $k$ propagate through and reshape the parasitic interaction network. 
\begin{equation}
\mathrm{SI}(k) = \sum_{(i,j)\in E^{+}} S_{k \rightarrow (i,j)},
\label{eq:sik}
\end{equation}

\emph{Algorithm~\ref{alg:csi_reconstruction}} returns the complete SI profile across all nodes, along with the deviation tensors and per-edge sensitivity matrices (\textit{lines~15–16}). Together, these outputs enable a precise and actionable characterization of the device’s dynamical vulnerability to local parameter fluctuations, allowing us to accurately pinpoint geometrically exposed nodes whose behavior can compromise system integrity. This insight directly informs layout refinement, coupler engineering, and compiler-level isolation and routing strategies.

\section{Evaluation} \label{sec:eval}

\subsection{Experimental Setup} \label{subsec:expsetup}

\noindent\underline{\textbf{Tools and implementation}}:
For our experiments, we utilized Qiskit Metal (v0.1.0) for layout generation, Gmsh (v4.15) for geometric meshing, and ElmerSolver (v9.0) for finite-element capacitance extraction~\cite{Qiskit_Metal, gmesh, ELMER}.
% The reconstructed interaction parameters and effective coupling graphs were obtained through our EPAR post-processing pipeline.
Time-domain simulations of the resulting multi-mode Hamiltonians were carried out using Qiskit Dynamics (v0.5), executed with a Duffing-model approximation~\cite{qiskit_dynamics_2023, koch2007charge}.

\noindent\underline{\textbf{Benchmark layouts and pulse workloads}}:
We evaluate EPAR using two complementary benchmark families. \textbf{First}, Sections~4.2 and~4.3 analyze a set of nine physically realistic Q$_i$--coupler--Q$_{i+1}$ module layouts representative of typical transmon–coupler design variations, providing the controlled diversity needed for LTD and SI studies. \textbf{Second}, Section~4.4 examines a representative five-qubit linear chain (Q$_0$--Q$_4$) with its associated couplers and drive lines to assess system-level implications of EPAR metrics for routing, scheduling, and compiler decisions.

To capture pulse-driven effects across all analyses, each benchmark family is simulated under multiple representative control-pulse families, including cross-resonance variants and coupler-modulation waveforms with different durations and amplitudes.

% \begin{figure}[t]
% % \vspace{-4mm}
% \centering
%   \includegraphics[width=0.8\linewidth]{figures/q_tc_circuit.pdf}
%   % \vspace{-5mm}
%   \caption{.}
%   \label{fig:q_tc_circuit}
%   \vspace{-5mm}
% \end{figure}

% as our measure of \emph{dynamic robustness}.
% This quantity appears directly in the LTD scatter plot and as the heatmap signal in the SI
% four-panel figure, linking geometric distortion and sensitivity to the resulting operational
% behavior.
% This one-dimensional sweep enables a clean mapping from geometry to both static distortion (LTD) and dynamic behavior (SI and relative infidelity) across all pulse families evaluated in this section.

\noindent\underline{\textbf{Evaluation metrics}}:
We evaluate the impact of layout variations using three core metrics.  
\textit{(i) Logical Topology Distortion (LTD)}: a static measure that captures how strongly the geometry deviates from the intended interaction pattern.  
\textit{(ii) Sensitivity Index (SI)}: a directional metric that quantifies how local perturbations at a node propagate through the reconstructed interaction graph. 
\textit{(iii) Dynamic robustness}: measured using the \emph{relative infidelity}, defined as $1 - F_{\mathrm{rel}}$, where $F_{\mathrm{rel}}$ denotes the overlap fidelity between the final pulse-driven state of a given layout and that of a reference layout under the same control pulse~\cite{jozsa1994fidelity}. 
% For a given layout $\ell$ and pulse family $p$, let 
% $\ket{\psi_{\mathrm{ref}}^{(p)}(t_{\mathrm{gate}})}$ and 
% $\ket{\psi_{\ell}^{(p)}(t_{\mathrm{gate}})}$ denote the final states at the gate time $t_{\mathrm{gate}}$, we define
% \begin{equation}
%     F_{\mathrm{rel}}(\ell,p)
%     = \bigl|\braket{\psi_{\mathrm{ref}}^{(p)}(t_{\mathrm{gate}})}{\psi_{\ell}^{(p)}(t_{\mathrm{gate}})}\bigr|^{2}
% \end{equation}
We use relative infidelity because it directly measures how strongly a layout
distortion alters the pulse-driven evolution itself.
% \textit{(iii) Dynamic robustness}: the relative fidelity between the pulse-driven evolution of each layout and that of a baseline geometry, evaluated across multiple representative control-pulse families.

\begin{figure}[b]
\vspace{-4mm}
\captionsetup[subfigure]{justification=centering}
 \centering
 \begin{subfigure}[b]{0.49\linewidth}
   \includegraphics[width=\textwidth]{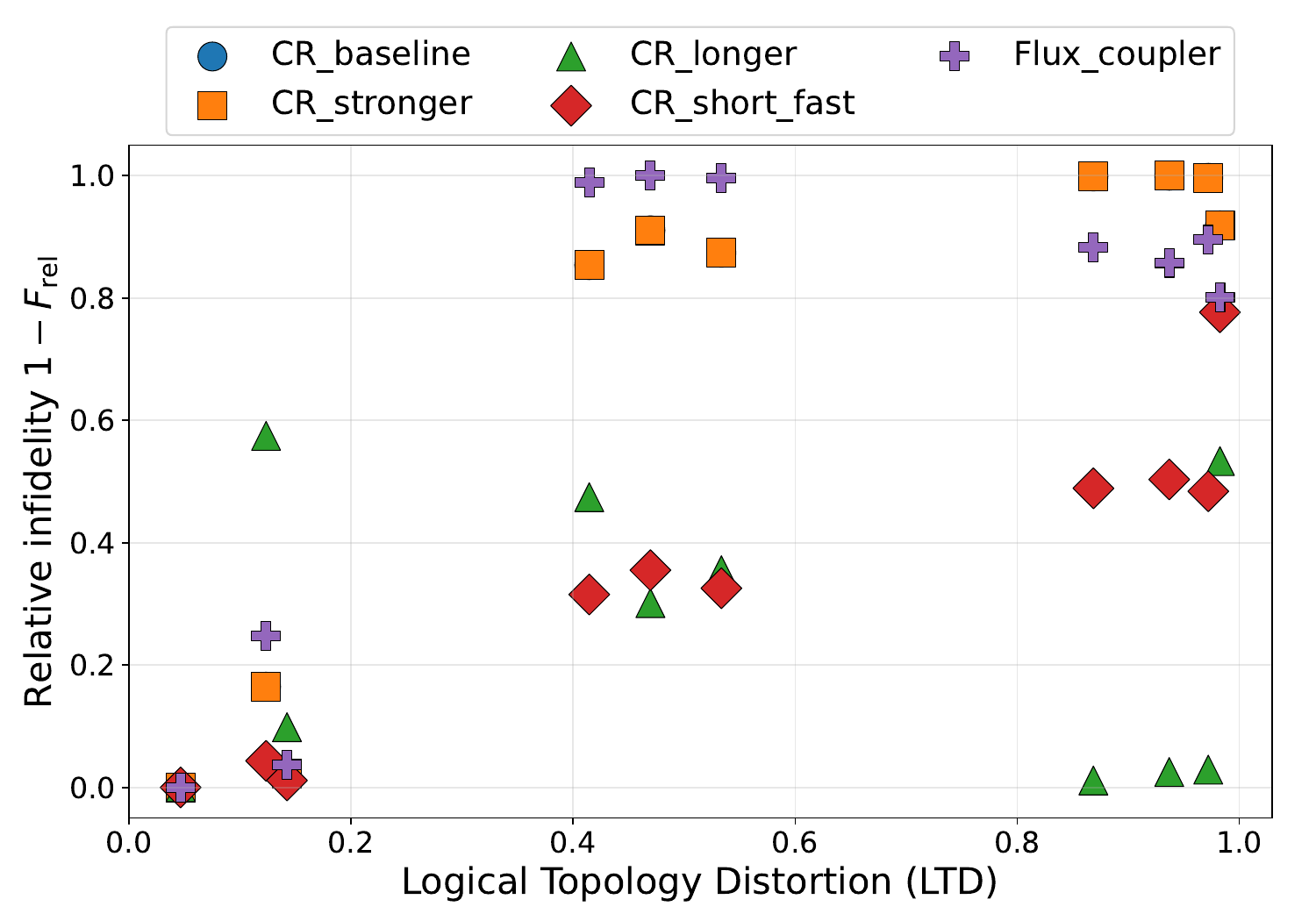}
   \caption{ Design Sweep A.}
 \label{subfig:cwidth}
 \end{subfigure}
 \begin{subfigure}[b]{0.49\linewidth}
  \includegraphics[width=\textwidth]{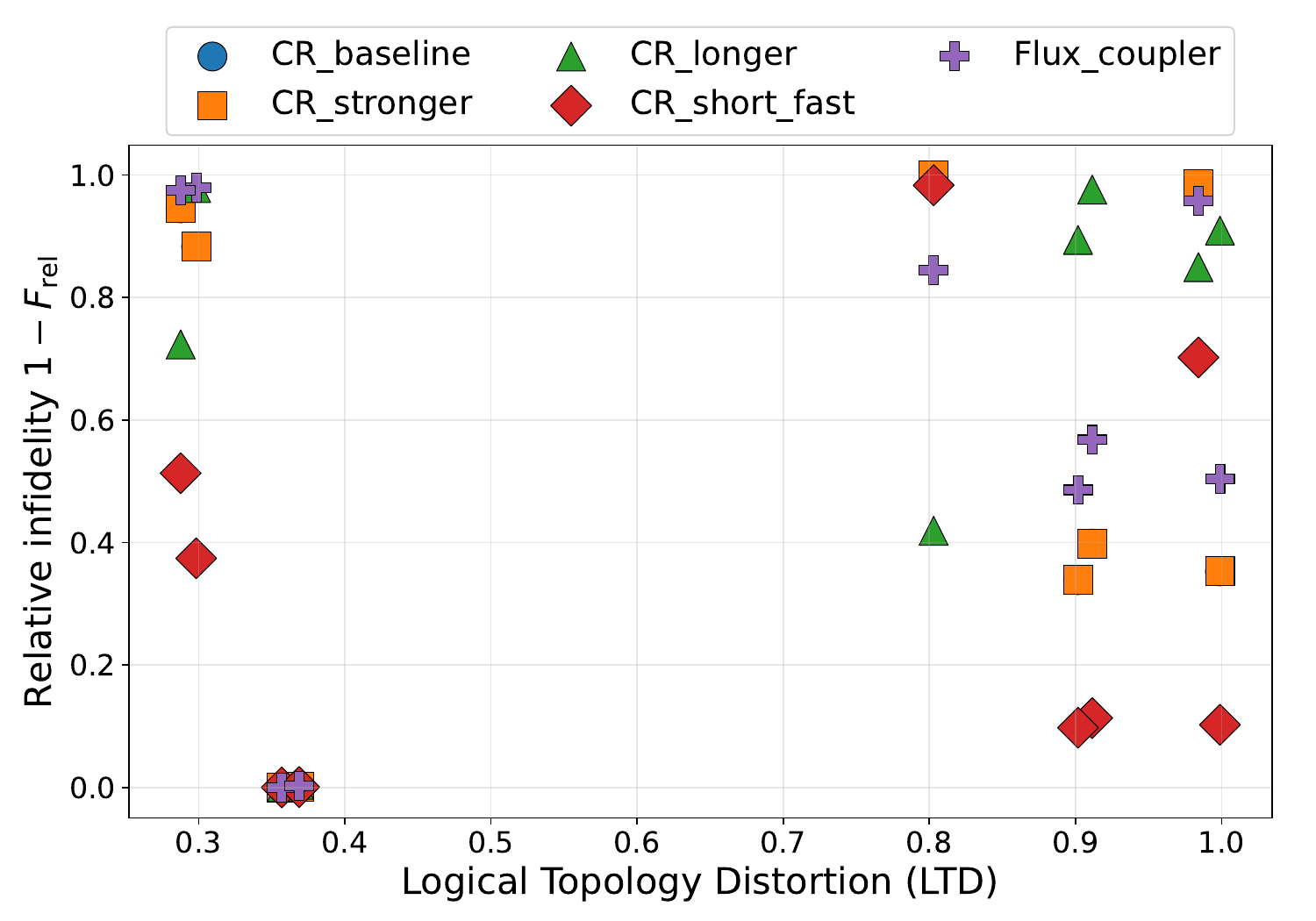}
 % \vspace{-8mm}
    \caption{Design Sweep B.}
    \label{subfig:aheight} 
 \end{subfigure}
 \caption{LTD exposes layout-induced distortion and identifies structurally weak edges across design sweeps.}
  % \vspace{-8mm}
\end{figure}

\subsection{Topology Distortion Across Designs}

Unlike raw capacitance or coupling values, LTD lets us determine whether the extracted Hamiltonian still supports a clean mediated interaction or has drifted into an unintended connectivity pattern. In Figures~\ref{subfig:cwidth} and~\ref{subfig:aheight}, we present two controlled design sweeps—Sweep~A (coupler width: 650\,nm--400\,nm) and Sweep~B (island height: 60\,µm--300\,µm)—each plotting LTD against relative infidelity for the resulting Q$_0$--C--Q$_1$ layouts.
By varying the coupler width, we sweep the geometry along the x-axis, and by varying the island height, we sweep geometry along the y-axis, covering the dominant geometric degrees of freedom and giving us two physically distinct yet controlled ways to probe how LTD responds to layout variation.

Using Figures~\ref{subfig:cwidth} and~\ref{subfig:aheight}, we show that LTD organizes the design space into distinct dynamical regimes, revealing when control pulses remain robust, become fragile, or fail entirely.
For low distortion (LTD~$<0.15$), all pulse families remain robust with low deviation $(1 - F_{\mathrm{rel}} \lesssim 10^{-3})$, showing that the extracted Hamiltonian still realizes the intended mediated interaction. In the intermediate range (LTD~$=0.3$--$0.6$), both sweeps show narrow geometry-dependent stability pockets—at very narrow and wide widths in Sweep~A and at 60\,µm and 200\,µm in Sweep~B (LTD~$\approx 0.36$ with very small infidelity)—surrounded by strong pulse-dependent fragility where sharper CR pulses amplify asymmetries and smoother pulses partially suppress them. For high distortion (LTD~$>0.8$), both sweeps collapse to large errors $(1 - F_{\mathrm{rel}} \approx 0.8$--$1.0)$ as the design effectively shifts into a different topology; in Sweep~B this includes island heights of 100\,µm and 250\,µm (LTD~$\sim 0.9$) and a failure point at 150\,µm where LTD reaches 0.984 and infidelity approaches unity.
\textit{Taken together, Figures~\ref{subfig:cwidth} and~\ref{subfig:aheight} show that LTD generalizes across both sweeps, with LTD$<0.15$ indicating robust operation, LTD$=0.3$--$0.6$ producing narrow design-dependent stability pockets with pulse-sensitive fragility, and LTD$>0.8$ yielding consistent failure with $(1-F_{\mathrm{rel}})\!\approx\!0.8$--$1.0$, providing actionable guidance for selecting robust designs and avoiding sensitive operating points.}

% Overall, LTD provides a \emph{system-level predictor} of when the control stack will remain stable, when it becomes pulse-sensitive, and when it fundamentally breaks down, offering actionable guidance for selecting coupler dimensions, mitigating parasitic asymmetries, and ensuring that device geometry remains within a regime where pulse protocols behave predictably.
\subsection{Sensitivity Across Control Regimes}
\begin{figure}[t]
% \vspace{-4mm}
\captionsetup[subfigure]{justification=centering}
 \centering
 \begin{subfigure}[b]{0.48\linewidth}
   \includegraphics[width=\textwidth]{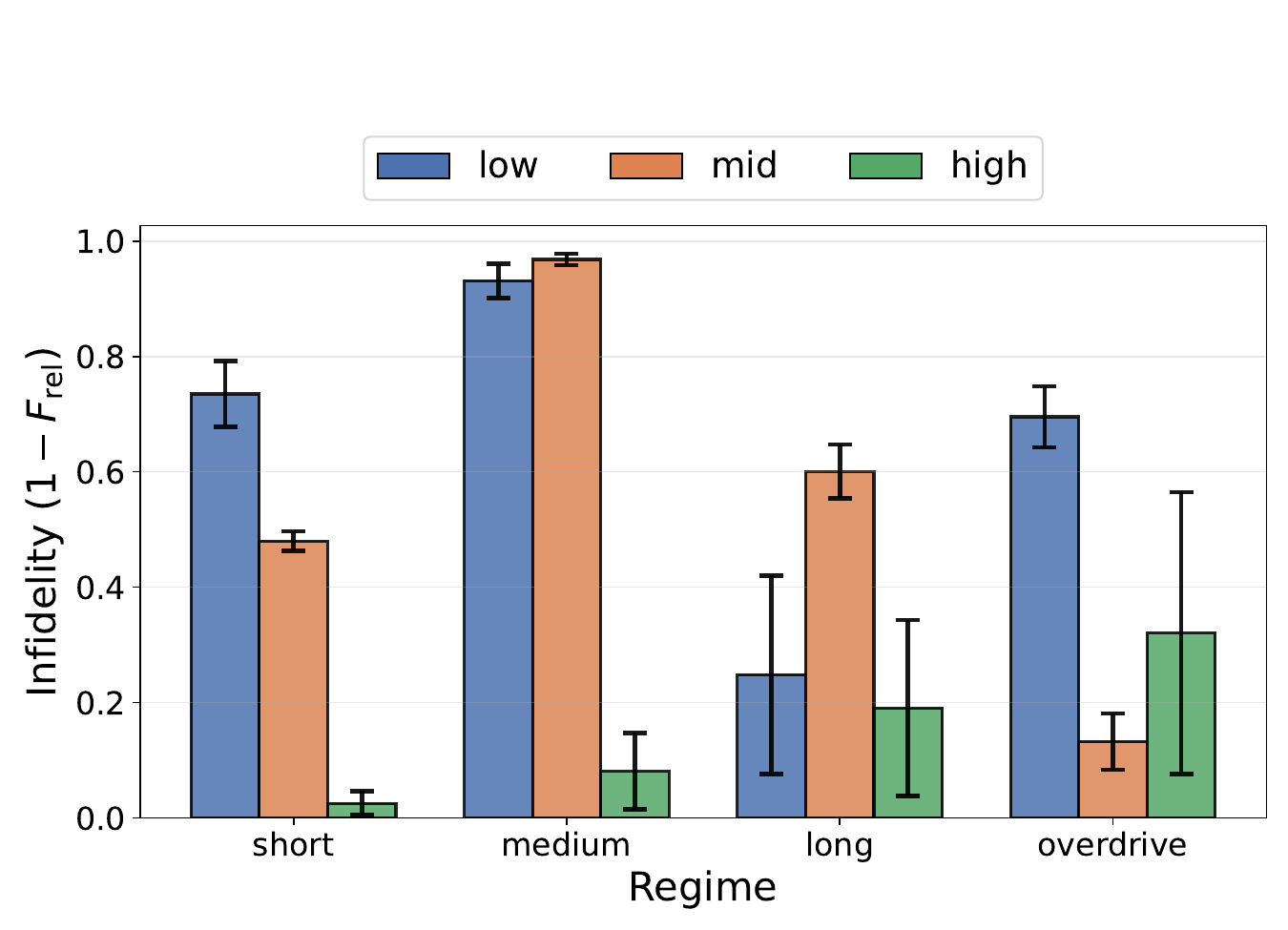}
   \caption{ Sensitivity–Fidelity Breakdown.}
 \label{subfig:si_bucket}
 \end{subfigure}
 \begin{subfigure}[b]{0.5\linewidth}
  \includegraphics[width=\textwidth]{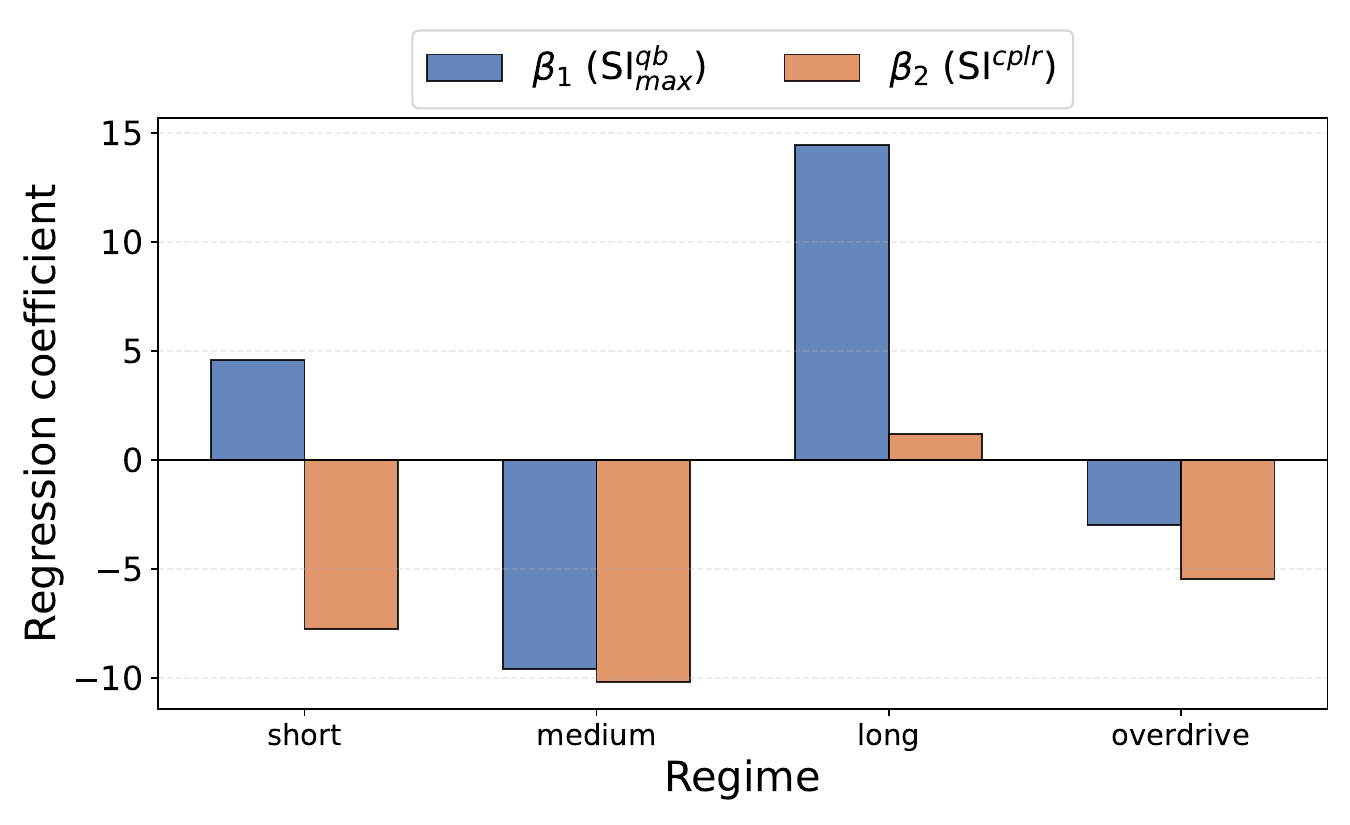}
 % \vspace{-8mm}
    \caption{Dominant Sensitivity Channels.}
    \label{subfig:si_regression} 
 \end{subfigure}
 \vspace{-2mm}
  \caption{SI uncovers regime-dependent fragility, uncovering the sensitivity pathways responsible for pulse-driven error.}
  \vspace{-5mm}
\end{figure}

Figures~\ref{subfig:si_bucket} and~\ref{subfig:si_regression} show how the Sensitivity Index (SI) governs fidelity degradation across regimes. 
Both SI plots use four pulse regimes—short ($\approx$20--30\,ns), medium ($\approx$40--60\,ns), long ($\approx$80--120\,ns), and overdrive ($\approx$150--220\,ns).
% each grouping pulse families (e.g., $CR_{\text{baseline}}$, $CR_{\text{stronger}}$, $CR_{\text{longer}}$, $\text{Flux}_{\text{cplr}}$) with similar temporal and amplitude profiles.
Using Figure~\ref{subfig:si_bucket}, we show that although the underlying SI values fall within a relatively tight numerical range across our geometry sweep, \emph{bucketing} layouts by SI reveals clear regime-dependent differences in dynamical stability. In the \textbf{short} regime, low-SI layouts exhibit the highest degradation ($\approx 0.73$), mid-SI layouts show moderate error ($\approx 0.48$), and high-SI layouts perform best ($\approx 0.03$). In the \textbf{medium} regime, low- and mid-SI layouts collapse toward near-total failure ($\approx 0.93$--$0.96$), whereas high-SI layouts remain substantially more stable ($\approx 0.08$). In the \textbf{long} regime, we observe a broadened spread: mid-SI layouts degrade most ($\approx 0.60$), low-SI layouts improve ($\approx 0.25$), and high-SI layouts maintain lower error ($\approx 0.19$). Under \textbf{overdrive}, the ordering reverses again—low-SI layouts degrade strongly ($\approx 0.69$), mid-SI layouts remain low-error ($\approx 0.13$), and high-SI layouts fall in between ($\approx 0.33$).
As each SI band aggregates multiple geometries (e.g., different coupler widths and island spacings), the error bars in Figure~\ref{subfig:si_bucket} capture \emph{intra-band geometric variation}, illustrating that layouts with similar SI can respond differently depending on the pulse regime. As observed, SI is not a monotonic predictor of performance: its impact is strongly regime-dependent, with different pulse families activating or suppressing distinct parasitic pathways.

To further explain the regime-dependent behavior, we use Figure~\ref{subfig:si_regression} to perform a regression analysis that isolates how qubit-side and coupler-side sensitivity channels contribute to error formation. In the \textbf{short} regime, the coefficients have opposite signs ($\beta_1\!\approx\!+4.7$, $\beta_2\!\approx\!-7.7$), with qubit-side sensitivity slightly increasing error and coupler-side sensitivity suppressing it, as fast pulses probe only the most immediate parasitic pathways. In the \textbf{medium} regime, both coefficients become strongly negative ($\beta_1\!\approx\!-9.5$, $\beta_2\!\approx\!-10.0$), showing that even modest sensitivity increases drive layouts toward near-total failure due to broadband excitation. In the \textbf{long} regime, this trend reverses: qubit-side sensitivity becomes the dominant positive predictor of degradation ($\beta_1\!\approx\!+14.5$) while coupler sensitivity weakens ($\beta_2\!\approx\!+1.2$), revealing long-timescale qubit-mediated error accumulation. Under \textbf{overdrive}, both coefficients again turn negative ($\beta_1\!\approx\!-3.0$, $\beta_2\!\approx\!-5.3$), indicating that strong pulses activate global breakdown channels in which both sensitivity pathways amplify parasitic interactions.
\textit{Overall, SI exposes layout-induced vulnerabilities that neither LTD nor calibrated two-qubit error rates reveal, supporting regime-aware, geometry-aware scheduling and routing strategies for compilation. }
% beyond what existing compilation can provide.
% highlighting (i) \emph{regime-dependent sensitivity} (medium: qubit- and coupler-driven; long: qubit-dominated), (ii) \emph{non-monotonic degradation} where mid-range SI fails first, and (iii) practical \emph{operational thresholds} ($\mathrm{SI}<0.10$ stable, $0.10$--$0.15$ fragile, $>0.15$ high risk). These empirically derived ranges 
% support regime-aware, geometry-aware scheduling and routing strategies beyond what error-rate--only compilation can provide.
\vspace{-1mm}
\subsection{EPAR for Reliability-Aware Compilation}
Using Table~\ref{tab:compiler_epar}, we connect EPAR’s metrics to compiler-facing quantities, and evaluate industry-standard two-qubit error rates as the baseline. For this analysis, we evaluate a representative five-qubit linear chain (Q0--Q4), constructed from the same design library as our earlier Q0--C--Q1 benchmarks so that each edge reflects a physically consistent layout. All Avg values are computed over \textit{four} design-level geometric knobs, namely, our sweep—coupler width (\texttt{cwidth}), island spacing (\texttt{aheight}), coupler–qubit gap, and coupler arm length, to capture each edge’s typical behavior across realistic layout variations.
\textbf{Column one} (Edge ID) lists each coupler chain, while \textbf{column two} (Avg LTD) reports EPAR’s structural distortion. Next, \textbf{column three} (Avg $\varepsilon_{2Q}$) provides the baseline 2Q error, and \textbf{column four} (Worst-Case Infidelity) gives the maximum $(1-F_{\mathrm{rel}})$ across pulses. \textbf{Column five} (Instability) measures the spread of relative infidelity, whereas \textbf{column six} (EPAR~$\leftrightarrow$~2Q Alignment) shows whether structure matches dynamic. Finally, \textbf{column seven} (Compiler-Level Decision) converts these metrics into routing and scheduling guidance.

\textbf{From the table}, we observe that despite having similar calibrated error behavior, the edges differ dramatically in structural and dynamical robustness. Q0--Q1 shows very high LTD but \emph{minimal} instability ($0.3\%$), identifying it as a structurally poor but predictably behaving edge—useful only as a spectator. Q1--Q2 and Q2--Q3 exhibit moderate LTD but \emph{large} instability ($11$--$11.4\%$), indicating geometry-amplified fragility that a compiler would never detect from average two-qubit error alone. Q3--Q4 is the most unstable ($21\%$ instability, worst-case infidelity $23.3\%$), despite average performance that could appear acceptable in a static calibration snapshot.
Overall, EPAR exposes structural and dynamical distinctions that error rates miss, with LTD providing structural visibility and SI adding regime-aware dynamics. Together, these insights enable \textbf{robustness-aware compilation} grounded in device behavior rather than single-point error metrics.
% \vspace{-4mm}
% SSSSSSSSSSSSSSSSssss
\begin{table}[t]
\centering
\caption{EPAR accurately predicts runtime instability.}  
\vspace{-3mm}
\label{tab:compiler_epar}

\resizebox{\linewidth}{!}{%
\Large
\begin{tabular}{ccccccc}
\hline
\textbf{Edge} &
  \textbf{Avg LTD} &
  \textbf{Avg $\varepsilon_{2Q}$ (\%)} &
  \textbf{Worst-case Infidelity (\%)} &
  \textbf{Instability (\%)} &
  \textbf{EPAR $\leftrightarrow$ 2Q} &
  \textbf{Compiler-level Decision} \\ \hline
Q0--Q1 & 4.097 & 5.64  & 5.78  & 0.30  & $\checkmark$ & \begin{tabular}[c]{@{}c@{}}Structural hot spot; \\ avoid entangling use.\end{tabular}              \\
Q1--Q2 & 0.720 & 10.69 & 13.75 & 11.20 & $\checkmark$ & \begin{tabular}[c]{@{}c@{}}Geometry-sensitive; \\ avoid long-lived logical paths.\end{tabular}     \\
Q2--Q3 & 0.833 & 10.25 & 15.31 & 11.36 & $\checkmark$ & \begin{tabular}[c]{@{}c@{}}Marginal edge; \\ use only under routing pressure.\end{tabular}         \\
Q3--Q4 & 1.071 & 14.39 & 23.34 & 21.18 & $\checkmark$ & \begin{tabular}[c]{@{}c@{}}Highly geometry-sensitive; \\ restrict parallel schedules.\end{tabular} \\ \hline
\end{tabular}%
} % end resizebox
\vspace{-3mm}
\end{table}
\vspace{-3mm}

\section{Related Work}
While superconducting processors are rapidly scaling in size and performance, existing calibration of the devices offers little visibility into layout-dependent electromagnetic effects. Despite evidence that layout-driven effects alter multi-qubit dynamics, their impact on system-level execution metrics remains absent from architectural tools~\cite{sung2021realization}. Work in quantum error correction and mitigation (e.g., Qiskit QEC, Stim) improves robustness but continues to operate on abstracted or empirical noise models rather than geometry-derived interactions. While tools like HFSS, and Qiskit Metal offer detailed component-level modeling, and EM–circuit co-models expose particular parasitic or crosstalk mechanisms, these analyses do not capture how geometry-driven interactions integrate, accumulate, or propagate through the system’s effective topology~\cite{hfss, sonnet, Qiskit_Metal}. Cryogenic-CMOS control architectures address wiring and integration challenges but remain orthogonal to understanding how physical geometry shapes computational behavior~\cite{cryo}. EPAR fills this gap by providing the first end-to-end, physics-grounded framework that derives the effective coupling topology directly from the layout and quantifies how layout-driven distortions propagate into mapping, scheduling, and routing decisions through new metrics such as LTD and SI.
\vspace{-3mm}

\section{Conclusion}

% In this paper, we presented EPAR, a geometry-aware analysis framework that links electromagnetic layout features to the effective interactions that govern superconducting qubit behavior. 
% We introduced LTD, which captures how much the extracted connectivity deviates from the intended topology, and showed that layouts with LTD $< 0.15$ maintain high dynamical accuracy ($1 - F_{\mathrm{rel}}< 10^{-3}$), while moderate distortion (0.3--0.6) produces pulse-dependent errors and severe distortion (LTD $> 0.8$) drives infidelity toward unity. We also introduced SI, which quantifies how local perturbations propagate through the system: SI $< 0.1$ corresponds to stable control, 0.15--0.35 marks emerging fragility, and SI $> 0.25$ aligns with rapidly rising errors and collapsing robustness margins ($\pm 30\% \rightarrow \pm 1\%$). Together, these metrics allow EPAR to identify reliable geometries, diagnose distortion-amplifying layouts, and provide actionable guidance for pulse design and compiler routing.
In this paper, we presented EPAR, a geometry-aware analysis framework that links electromagnetic layout features to the effective interactions that govern superconducting qubit behavior. We introduced LTD, which captures how much the extracted connectivity deviates from the intended topology, and showed that layouts with LTD $<0.15$ maintain high dynamical accuracy ($1-F_{\mathrm{rel}}!\lesssim!10^{-3}$), while moderate distortion (0.3–0.6) produces geometry- and regime-dependent pulse errors, and severe distortion (LTD $>0.8$) drives infidelity toward unity. We also introduced SI, which quantifies how local perturbations propagate through the system: SI bands separate layouts that remain stable across regimes from those that fail under medium and overdrive pulses, revealing up to 10$\times$ robustness differences among geometries with similar calibrated two-qubit errors. Together, these metrics allow EPAR to identify reliable designs, diagnose distortion-amplifying layouts, and provide actionable guidance that complements existing compiler heuristics.

\balance

% \clearpage

% \input{sections/ethical_consideration}

%\balance
\bibliographystyle{IEEEtran}
\bibliography{refs}

\end{document}